\documentstyle[12pt]{article}
\title{Nucleon Axial Matrix Elements}
\author{Barry R. Holstein\\
Department of Physics and Astronomy\\
University of Massachusetts\\
Amherst, MA  01002   USA\\
and\\
Institut f\"{u}r Kernphysik\\
Forschungszentrum J\"{u}lich\\
D-52425 J\"{u}lich, Germany}
\begin{document}
\begin{titlepage}
\maketitle
\begin{abstract}
Current issues associated with nucleon axial matrix elements are studied,
including
the Goldberger-Treiman discrepancy, the induced pseudoscalar, and SU(3) chiral
perturbation theory.
\end{abstract}
\end{titlepage}
\section{Introduction}
I have been given the task of speaking about the nucleon axial matrix elements.
 In
comparison with the many exciting things being discussed at this meeting this
may
seem rather prosaic.  However, I will try to convince you otherwise by
discussing
issues associated with the Goldberger-Treiman discrepancy, recent and future
measurements of the induced pseudoscalar, and the renormalizations of
the axial couplings within SU(3) chiral perturbation theory.

\section{The Goldberger-Treimen Discrepancy OR Time Dependence of Fundamental
Constants}

Many years ago Dirac noticed that the ratio of electrical to gravitational
forces
between a pair of electrons was equal to $\alpha/(Gm_e^2)\sim 10^{40}$.  In
asking
himself how such an enormous dimensionless number could arise he noticed that
$10^{40}$ is also the age of the universe measured in fundamental units of
time---{\it i.e.} the time it takes light to traverse an elementary
particle---$10^{17}$ sec./ $10^{-23}$ sec.!  He then asked if it were possible
that
the electrical to gravitational ratio might change as the universe evolved.  It
turns
out on further analysis that this is extremely unlikely---the consequences of
even
relatively small changes to either the fine structure or gravitational
constants
turn out to be significant (the anthropic principle),\cite{anth}
but I am prepared today to
point out an arena where the changes in a fundamental coupling have been major,
and
they have occured within a generation----the nucleon axial coupling.  Below I
give
a list of values which I have gathered from various sources, and it is clear
that
there has been a seven percent increase in $g_A$ within a decade!
\begin{equation}
\begin{array}{ll}
1959:  g_A/g_V= 1.17\pm 0.02\cite{av1}&
1965:  g_A/g_V=1.18\pm 0.02\cite{av2}\\
1967:  g_A/g_V=1.24\pm 0.01\cite{av3}&
1969:  g_A/g_V= 1.26\pm 0.02\cite{av4}
\end{array}
\end{equation}
In fact this number continues to increase---the latest published experiments
from
Grenoble give $g_A/g_V=1.266\pm 0.004$,\cite{Gren}
which is the value I shall employ in this
note.

Now my facetious discussion in the previous paragraph is at one level amusing,
but
at another has some important ramifications when considered in terms of the
Goldberger-Treiman relation\cite{GT}
\begin{equation}
M_Ng_A(0)=F_\pi g_{\pi NN}(0)\label{eq:aa}
\end{equation}
which is required by chiral invariance and hence by QCD.  Now I have indicated
in
Eq. \ref{eq:aa} that the axial coupling $g_A$ and the pion-nucleon coupling
constant $g_{\pi NN}$ are {\it both} to be evaluated at zero momentum transfer.
However while the former is the ("time-dependent") number quoted above, the
latter
is {\it not} a physical quantity.  What {\it is} directly measurable is the
pion-nucleon
coupling evaluated at the pion mass-squared---$g_{\pi NN}(m_\pi^2)$, so it is
useful to examine the Goldberger-Treiman {\it discrepancy}
\begin{equation}
\Delta_\pi=1-{Mg_A(0)\over F_\pi g_{\pi NN}(m_\pi^2)}
=1-{g_{\pi NN}(0)\over g_{\pi NN}(m_\pi^2)}
\end{equation}
In the venerable text Bjorken and Drell this number is described as being
less than 0.1, but I want to argue that it must be {\it much} less.  Indeed,
while
$g_{\pi NN}(0)$ is not an observable, one can show that in reasonable models
such
as the linear sigma model or via correlated $\pi -\rho$ exchange one should
expect
$\Delta_\pi\simeq 0.02$.  However, there is another interesting approach via
the
so-called Dashen-Weinstein theorem,\cite{dwt}
which uses the fact that while there does not
exist a prediction for $\Delta_\pi$ in chiral SU(2), since it is given in terms
of
an a priori unknown counterterm, in SU(3) this
quantity is given in terms of a sum of quark
masses times an SU(3) octet operator.  Thus a relation exists---the
Dashen--Weinstein
theorem---{\it between} corresponding kaon and pion quantities\cite{dom,stern}
\begin{equation}
\Delta_\pi={\sqrt{3}\over 2}{F_K\over F_\pi}{m_u+m_d\over m_u+m_s}\left(
{g_{\Lambda KN}\over g_{\pi NN}}\Delta^\Lambda_K-{1\over \sqrt{6}}
{g_{\Sigma KN}\over g_{\pi NN}}\Delta^\Sigma_K\right)
\end{equation}
where
\begin{equation}
\Delta_K=1-{(M_N+M_{\Sigma,\Lambda})g_A(0)\over \sqrt{2}F_Kg(m_K^2)}=\left\{
\begin{array}{ll}
0.32&\Lambda\\
-0.05&\Sigma
\end{array}\right.
\end{equation}
where I have used the values\cite{pdg,mart}
\begin{equation}
\begin{array}{ccc}
&g(m_K^2)& g_A(0)\\
\Lambda & -13.5& -0.72\\
\Sigma & 4.3&0.34
\end{array}
\end{equation}
Then using $(m_u+m_d)/(m_u+m_s)=m_\pi^2/m_K^2$ we find $\Delta_\pi^{theo}\simeq
0.028.$
Besides changes in the size of the axial coupling, however,
the size of $F_\pi$ decreased by
1\% in 1990 when it was realized that previous evaluations had not included the
running of the weak coupling constant,\cite{h90}
and there has been continuous debate about
the size of $g_{\pi NN}(m_\pi^2)$, with current analyses favoring either
the Karsruhe value 13.4\cite{kr} or the VPI number 13.05.\cite{vpi}  Thus we
find
\begin{eqnarray}
g_{\pi NN}=13.4\longrightarrow \Delta_\pi=4.1\% ,\quad{\rm or} \quad
m_s/\hat{m}\approx 48
\nonumber\\
g_{\pi NN}=13.05\longrightarrow \Delta_\pi=1.5\% ,\quad{\rm or} \quad
m_s/\hat{m}
\approx 17
\end{eqnarray}
so that if the low value of $g_{\pi NN}$ is confirmed, the Goldberger-Treiman
discrepancy would strongly favor the conventional $\chi pt$ picture
($m_s/\hat{m}=25$)
over its generalized version, which predicts $m_s/\hat{m}<25$.\cite{stern}

\section{The Induced Pseudoscalar}

The axial matrix element of the nucleon consists in general of two pieces, the
usual
axial coupling and the induced pseudoscalar\footnote{In general one could also
allow
an axial tensor coupling, but this "second class current" is disallowed by
G-invariance.}
\begin{equation}
<p(p')|A_\mu|n(p)>=\bar{u}(p')(g_A(q^2)\gamma_\mu\gamma_5+g_P(q^2)
{q_\mu\over 2M}\gamma_5)
u(p)
\end{equation}
and chiral considerations require that this new piece is dominated by its pion
pole
contribution
\begin{equation}
g_P(q^2)={4MF_\pi\over m_\pi^2-q^2}g_{\pi NN}(q^2)\simeq {4MF_\pi\over
m_\pi^2-q^2}
g_{\pi NN}(m_\pi^2)-{2M^2\over 3}g_A(0)r_A^2
\end{equation}
where $r_A$ is the axial radius.  This result is generally used in the
combination
\begin{equation}
r_P={m_\mu\over 2Mg_A(0)}g_P(q^2=-0.9m_\mu^2)=6.7
\end{equation}
relevant for muon capture.  This is the standard approach and is used
because the contraction of the four-vector
$q_\mu$ with the lepton tensor results in a factor of the lepton mass
accompanied
by the nucleon matrix element of $\gamma_5$, which brings in an additional
suppression
$|\vec{q}|/2M$, meaning that despite the extraordinary precision of modern
nuclear
beta decay experiments, any effects from $g_P$ arise only at ${\cal O}[
r_Pm_e^2/(2Mm_\mu)]\sim 10^{-5}$!.  On the other hand in muon capture this
factor
becomes $r_Pm_\mu/2M$, which means that the pseudosclar contributes at the same
order as weak magnetism and becomes in principle measurable.  The
one problem here is that typically one has only a single number---the capture
rate---to
work with so that in order to extract the desired value of $r_P$ one must make
three reasonable, but still model-dependent, assumptions---i)
the validity of CVC in order to extract $f_V(q^2),f_M(q^2)$ from electron;
scattering data; ii) the validity of the impulse approximation to evaluate
$g_A(q^2)$;
and iii) the assumption of G-invariance to rule out the presence of second
class
currents.  Using these assumptions one finds the experimental values
\begin{equation}
r_P=\left\{\begin{array}{ll}
6.5\pm 2.4& H\cite{hr}\\
6.9\pm 0.2& {}^3He\cite{her}\\
9.0\pm 1.7& {}^{12}C\cite{cr}
\end{array}
\right.
\end{equation}
which are in agreement with the chiral expectations.  It should be noted that
the
extraordinary precision associated with the $^3He$ number is allowed because of
a
spectacular new PSI experiment which measured the capture rate to 3\%
\begin{equation}
\Gamma_\mu({}^3He)=1496\pm 4 sec.^{-1}
\end{equation}
In order to eliminate some of this model dependence, there are additional
approaches
which have been and which are being pursued
\begin{itemize}
\item [i)] Radiative muon capture on hydrogen:  This is the approach
which has received the most recent attention, because the
result\cite{rcap}
\begin{equation}
r_P=9.8\pm 0.7\pm 0.3
\end{equation}
is at variance with the chiral prediction at the $3\sigma$ level.  Now this
TRIUMF
measurement is extraordinarily difficult both because of the tiny $10^{-8}$
branching fraction compared to ordinary capture and because of the presence of
many possible experimental backgrounds.  However, it has the advantage that at
the
maximum photon energy $k_{max}=100 MeV$ the momentum transfer is
$q_{max}=m_\mu^2$
compared to the value $q_\mu^2=-0.9m_\mu^2$ which obtains in the ordinary muon
capture case.  Integrated over the photon spectrum this leads to an enhancement
of about a factor of three for pseudoscalar effects in RMC over those in
OMC.\cite{rth}
Clearly this is an experiment that should be repeated.

\item [ii)] Threshold pion photoproduction on hydrogen: This
might seem a strange place to study
nucleon {\it axial} matrix elements, but this is possible because of the PCAC
relation\cite{pcac}
\begin{equation}
<\pi^+n_{p'}|V^{em}_\mu|p_p>\stackrel{q\rightarrow 0}{\longrightarrow}
{-i\over \sqrt{2}F_\pi}<n_{p'}|A_\mu^-|p_p>
\end{equation}
The variation in $q^2$ which is allowed by the use of electroproduction rather
than photoproduction permits a check of the $q^2$-variation of both axial
matrix
elements.  A recent Saclay experiment produced in this way a measurement of the
axial radius $r_A$ which was in good agreement with parallel neutrino
scattering
measurements, when a small chiral symmetry offset is included, but more
importantly
for our case for the first time a measurement was made of the {\it shape} of
$g_P(q^2)$ which was in good agreement with the pion pole dominance
assumption.\cite{ppol}

\item [iii)] Correlations in polarized muon capture on ${}^3He$:
The final method which is being pursued at present goes back to an old idea to
measure the correlation of the final neutrino direction with initial state
polarization in the case that the muon and target are polarized.\cite{pcor}
Of course, the
muon is almost completely longitudinally polarized at the time of its capture,
but
unless the target too is polarized this polarization for the most part lost as
the muon cascades down through the various atomic levels before finally
reaching
the ground state--1S--level from which it is captured.  In the general case,
when
one has muon polarization $P'\hat{n}$ and (spin 1/2) target polarization
$P\hat{n}$
the decay distribution is found to be of the form
\begin{equation}
{d^2\Gamma_\mu\over d\Omega_{\hat{k}}}=A-2PP'B-{1\over
2}(P+P')C\hat{n}\cdot\hat{k}
+2PP'D\left[(\hat{n}\cdot\hat{k})^2-{1\over 3}\right]
\end{equation}
where the structure functions $A,B,C,D$ are functions of the weak formfactors
$g_v,g_M,g_A,g_P$ whose specific forms can be found in the literature.  The
important feature for our case is that D has a strong dependence on $g_P$, and
thus measurement of the angular correlations allows one to pick out the induced
pseudoscalar.  This measurement was attempted unsuccessfully many years ago at
LAMPF, but only recently has the ability to polarize ${^3He}$ at high levels
given hope that this experiment can actually be carried out.  Preliminary
results
at TRIUMF are encouraging, but the precision is not yet at a level where
anything
definitive can be said.\cite{trfp}
\end{itemize}

\section{Axial Couplings and SU(3)}

The existence of semileptonic hyperon decays such as
$\Lambda,n\rightarrow pe^-\bar{\nu}_e,\Sigma^-,\Xi^-\rightarrow \Lambda
e^-\bar{\nu}_e,
\Xi^-\rightarrow \Sigma^0e^-\bar{\nu}_e,$ etc. allows the probing of axial
matrix
elements in SU(3).  Indeed it has long been known that to leading order in
chiral
symmetry one can describe such decays in terms of simple f,d parameters, {\it
e.g.}
\begin{equation}
\begin{array}{ll}
g_V^{pn}=f_V& g_A^{np}=f_A+d_A\\
g_V^{p\Lambda}=-\sqrt{3\over 2}f_V&g_A^{p\Lambda}=-\sqrt{1\over 6}(3f_A+d_A)\\
g_V^{n\Sigma^-}=-f_V&g_A^{n\Sigma^-}=-f_A+d_A
\end{array}
\end{equation}
This type of fit yields remarkably good results---$\chi^2_{d.o.f}\approx 8.5$
for ten
degrees of freedom, when small $\leq 5\%$
quark model symmetry breaking effects are added.\cite{dhk}  One
can try to do even better by including chiral loops using heavy baryon chiral
perturbation theory.  At one loop one finds results\cite{bsw}
\begin{equation}
g_A^{ij}=(f_A,d_A)^{ij}+\sum_m\beta_m^{ij}m_m^2\ln{m_m^2\over \mu^2}
\end{equation}
However, this inclusion of supposedly {\it model-independent} corrections
brings in
modifications to the axial couplings at the level of 30-50\% which results in a
vastly increased $\chi^2$.  Of course, one can restore experimental agreement
by
the addition of appropriately chosen higher order counterterms, but then one
worries
about the convergence of the chiral expansion and is certainly justified in
asking what is going on.  Our answer is that this simple chiral picture
omits an important piece of physics, which is finite hadronic size.\cite{dhb}
The simple
chiral expansion assumes (at lowest order) propagation of mesons between {\it
point}
baryons, while in the real world any such propagation takes place between
objects
about a fermi or so in size.  That means that only the {\it long}-distance
component of the meson loop is really model-independent and to be trusted.  One
can
eliminate such short distance components by use of a cutoff regularization with
scale $\sim$300 MeV $\leq\Lambda\leq\sim$ 600 MeV of order inverse baryon size
rather than the usual dimensional regularization which mixes both long and
short distance effects.\cite{dh} The specific form of the cutoff function is
unimportant, so for calculational
purposes it is useful to use a simple dipole.  The result is that the heavy
baryon integral responsible for loop corrections to axial couplings
\begin{equation}
I_{ij}(m^2)=\int{d^4k\over(2\pi)^4}{k_ik_j\over
(k_0-i\epsilon)^2(k^2-m^2+i\epsilon)}
={-i\delta_{ij}\over 16\pi^2}m^2\ln{m^2\over \mu^2}\label{eq:jj}
\end{equation}
is replaced by
\begin{equation}
\tilde{I}_{ij}=\int{d^4k\over (2\pi)^4}{k_ik_j\over
(k_0-i\epsilon)^2(k^2-m^2+i\epsilon)}
\left({\Lambda^2\over \Lambda^2-k^2}\right)^2={-i\delta_{ij}\over
16\pi^2}J(m^2)
\end{equation}
where
\begin{equation}
J(m^2)={\Lambda^4\over (\Lambda^2-m^2)^2}m^2\ln{m^2\over
\Lambda^2}+{\Lambda^4\over
\Lambda^2-m^2}
\end{equation}
We see then that unlike Eq. \ref{eq:jj} which emphasizes heavy meson
(short-distance)
propagation over that of light mesons, in the large mass limit
\begin{equation}
J(m^2)\stackrel{m^2>>\Lambda^2}{\longrightarrow}{\Lambda^4\over m^2}
\ln{m^2\over \Lambda^2}\rightarrow 0
\end{equation}
On the other hand in the large cutoff limit we have
\begin{equation}
J(m^2)\stackrel{\Lambda^2>>m^2}{\longrightarrow}\Lambda^2+m^2\ln{m^2\over
\Lambda^2}
\end{equation}
which reproduces the usual dimensional regularization result plus a quadratic
term
in $\Lambda$.  That this latter piece does not destroy the chiral invariance
can
be seen from the feature that it can be absorbed in a renormalization of the
basic
couplings\cite{luty}
\begin{eqnarray}
d_A^r&=&d_A^{(0)}-{3\over 2}d_A(3d_A^2+5f_A^2+1){\Lambda^2\over 16\pi^2F_\pi^2}
\nonumber\\
f_A^r&=&f_A^{(0)}-{1\over 6}f_A(25d_A^2+63f_A^2+9){\Lambda^2\over
16\pi^2F_\pi^2}
\end{eqnarray}
However, in this procedure with reasonable values of the cutoff, the SU(3)
chiral expansion is now under control, as can be seen in Table 1.

\begin{table}
\begin{center}
\begin{tabular}{c|c|c|c|c|c}
  &dim.&$\Lambda$=300&$\Lambda$=400&$\Lambda$=500&$\Lambda$=600\\
\hline
$g_A(\bar{p}n)$&1.72&0.37&0.53&0.69&0.84\\
$g_A(\bar{p}\Lambda)$&-1.78&-0.34&-0.51&-0.67&-0.84\\
$g_A(\bar{\Lambda}\Sigma^-)$&1.17&0.23&0.34&0.45&0.56\\
$g_A(\bar{n}\Sigma^-)$&0.36&0.07&0.10&0.14&0.17\\
$g_A(\bar{\Lambda}\Xi^-)$&0.83&0.15&0.23&0.31&0.39\\
$g_A(\bar{\Sigma^0}\Xi^-)$&2.46&0.45&0.68&0.91&1.15
\end{tabular}
\caption{Given are the nonanalytic contribtions to $g_A$ for various
transitions in dimensional regularization and for various values of
the cutoff parameter $\Lambda$ in MeV.}
\end{center}
\end{table}
This brings such results into agreement with typical chiral bag calculations,
such
as the cloudy bag,\cite{clb}
and there is no longer any need to append large counterterm
contributions in higher orders.

\section{Conclusion}

We have above considered an old subject---that of nucleon axial matrix
elements---from the point of view of modern experiments.  I hope that I have
convinced
you that despite the age of the field, the new results in the areas
of Goldberger-Treiman
discrepancies, induced pseudoscalar measurements, and SU(3) chiral perturbative
studies promise continued interest even as we approach the millenium.

\begin{center}
{\bf Acknowlegement}
\end{center}
It is a pleasure to acknowledge support from the Alexander von
Humboldt Foundation and the hospitality
of Forschungszentrum J\"{u}lich.  This work was also supported by the National
Science Foundation.\\


\medskip
\begin{thebibliography}{99}
\bibitem{anth} For a recent slant and for further references see V. Agrawal et
al., hep-ph/9707380.
\bibitem{av1} A.N. Sosnovskii et al., Sov. Phys.  JETP {\bf 8}, 739 (1959).
\bibitem{av2} S.A. Adler, Phys. Rev. Lett. {\bf 14}, 1051 (1965).
\bibitem{av3} C.J. Christensen et al., Phys. Lett. {\bf B26}, 11 (1967); Phys.
Rev.
{\bf D5}, 1628 (1972).
\bibitem{av4} R.J. Blin-Stoyle, {\bf Fundamental Interactions and the Nucleus},
North-Holland, New York (1969).
\bibitem{Gren} K. Schreckenbach et al., Phys. Lett. {\bf B259}, 353 (1991).
\bibitem{dwt} R. Dashen and M. Weinstein, Phys. Rev. {\bf 188}, 2330 (1969).
\bibitem{dom} C.A. Dominguez, Riv. del Nuovo Cimento {\bf 8},\#6,1 (1985).
\bibitem{stern} N.H. Fuchs, H. Sazdjian, and J. Stern, Phys. Lett. {\bf B238},
380 (1990).
\bibitem{h90} B.R. Holstein, Phys. Lett. {\bf B244}, 83 (1990).
\bibitem{pdg} Particle Data Group, Phys Rev. {\bf D50}, 1173 (1995).
\bibitem{mart} H. Haberzettl et al., nucl-th/9804051.
\bibitem{kr} R. Koch and E. Pieterinin, Nucl. Phys. {\bf A336}, 331 (1980).
\bibitem{vpi} R. Arndt et al., Phys. Rev. Lett. {\bf 65}, 157 (1990).
\bibitem{GT} M.L. Goldberger and S.B. Treiman, Phys. Rev. {\bf 110}, 1478
(1958).
\bibitem{Bj} J.D. Bjorken and S.D. Drell, {\bf Relativistic Quantum Mechanics},
McGraw-Hill, New York (1964).
\bibitem{hr} G. Bardin et al., Phys. Lett. {\bf B104}, 320 (1981).
\bibitem{her} P. Ackerbauer et al., Phys. Lett. {\bf B417}, 224 (1998).
\bibitem{cr} V. Roesch et al., Phys. Rev. Lett. {\bf 46}, 1507 (1981).
\bibitem{rcap} G. Jonkmans et al., Phys. Rev. Lett. {\bf 77}, 4512 (1996).
\bibitem{rth} H.W. Fearing, Phys. Rev. {\bf C21}, 1951 (1980); D.S. Beder and
H.W. Fearing, Phys. Rev. {\bf D39}, 3493 (1989).
\bibitem{pcac} A.I Vainshtein and V.I. Zakharov, Nucl. Phys. {\bf B36} (1972);
V. Bernard, N. Kaiser, and U.-G. Meissner, Nucl. Phys, {\bf A607}, 379 (1996).
\bibitem{ppol} S. Choi et al., Phys. Rev. Lett. {\bf 71}, 3927 (1993).
\bibitem{pcor} See, {\it e.g.} B.R. Holstein, Phys. Rev. {\bf C3}, 1964 (1972).
\bibitem{trfp} W.J. Cummings et al., Proc. WEIN '95, ed. H. Ejiri, T.
Kishimoto, and
T. Sato, World Scientific, Singapore (1995), p. 381; G. Cates, private
communication.
\bibitem{dhk} J.F. Donoghue, B.R. Holstein, and S.W. Klimt, Phys. Rev.
{\bf D35}, 934 (1987]
\bibitem{bsw} J. Bijnens, H. Sonoda, and M.B. Wise, Nucl. Phys. {\bf B261},
185 (1985).
\bibitem{dhb} J.F. Donoghue, B.R. Holstein, and B. Borasoy, hep-ph/9804281.
\bibitem{dh} J.F. Donoghue and B.R. Holstein, hep-ph/9803312.
\bibitem{luty} M.A. Luty and M. White, Berkeley prepring LBL33993 (1993).
\bibitem{clb} See, {it e.g.} T. Yamaguchi et al., Nucl. Phys. {\bf A500}, 129
(1989);
K. Kubodera et al., Nucl. Phys. {\bf A439}, 695 (1985); S. Theberge et al.,
Phys.
Rev. {\bf D22}, 2838 (1980); A.W. Thomas, J. Phys. {\bf G7}, L283 (1981); R.E.
Stuckey and M.C. Birse, J. Phys. {\bf G23}, 29 (1997).






\end{thebibliography}
\end{document}